\documentclass{mn2e}\usepackage{graphicx}

\title{The X-ray, optical and radio evolution of the GRB~030329 afterglow and 
the associated SN2003dh}

\author[R. Willingale, J.P. Osborne, P.T. O'Brien...]
{R. Willingale,
J.P. Osborne,
P.T. O'Brien,
M.J. Ward, A. Levan and K.L. Page\\
Department of Physics and Astronomy, University 
of Leicester, University Road, Leicester LE1 7RH}

\date{}

\def\H0{{\rm ~km~s^{-1}~Mpc^{-1}}}

\def\la{\mathrel{\hbox{\rlap{\hbox{\lower4pt\hbox{$\sim$}}}{\raise2pt\hbox{$<$}}}}}
\def\ga{\mathrel{\hbox{\rlap{\hbox{\lower4pt\hbox{$\sim$}}}{\raise2pt\hbox{$>$}}}}}

\def\d25{D$_{25}$}

\def\.25{0.25 keV\thinspace}

\def\araa{ARA\&A}

\begin{document}

\maketitle

\begin{abstract}
Extensive X-ray, optical and radio observations of the bright 
afterglow of the Gamma Ray Burst GRB~030329 are used to construct the
multi-frequency evolution the event.
The data are fitted using the
standard fireball shock model to provide estimates of the initial energy, 
$\varepsilon = 6.8\times 10^{52}$ ergs~sr$^{-1}$, 
the density of the ambient medium,
$n_{0} = 1$ cm$^{-3}$, 
the electron and magnetic energy density fractions, 
$\epsilon_{e} = 0.24$ \& $\epsilon_{B} = 0.0017$,
the power law index of the relativistic electron spectrum,
${\rm p} = 2.25$, 
and the opening angle of the jet,
$\theta_{j} = 3$ degrees.
Deviations from the standard model seen in the optical and radio
are most likely attributable to the concurrent hypernova SN2003dh. 
Peaks at 0.23 and 1.7 days in the R-band are much brighter than expected
from a standard SN, and there is a large radio excess over the expected
afterglow flux for t$>$2 days.
No deviation from the best-fit afterglow model is seen in the X-ray
decline, indicating that the excess optical and radio flux from 1-5 days
arises from a later injection of slower electrons by the central engine.
\end{abstract}


\section{Introduction}

The trigger for GRB~030329 was made by HETE-II on 29 March 2003,
11:37:14.67 UT. Information promptly
disseminated via the GRB Coordinates Network
(GCN) by Vanderspek et al. (2003) prompted observations by a huge number of
telescopes providing coverage in the  X-ray, optical, IR, sub-millimetre and
radio wavebands. The R band magnitude at 1.5 hours was 12.6 making it the
brightest optical GRB afterglow seen at this epoch. The GRB lasted 25 seconds
and had a peak flux of $\sim7\times10^{-6}$ ergs cm$^{-2}$ s$^{-1}$ 
(30-400 keV) and a total fluence of $\sim1.2\times10^{-4}$ ergs cm$^{-2}$, 
placing it in the top 0.2\% of the observed fluence distribution.
The redshift was determined as z=0.1685 by Greiner et al. (2003)
and subsequently the isotropic energy released in the GRB has been
estimated to be $E_{iso}\approx9\times10^{51}$ ergs (30-400 keV)
(Hjorth et al., Price et al., Uemura et al. 2003).

\section{Afterglow spectrum of a fireball shock}

Here we summarise the continuum afterglow spectrum expected to arise from the
standard fireball shock model (M\`{e}sz\`{a}ros 2002 and references therein).
An initial energy $E_{52}=E/10^{52}$ ergs
is released very rapidly forming a fireball of
$e^{+}$ $e^{-}$ and gamma-rays which expands at extreme
relativistic velocity (Lorentz factor $\Gamma$) into a surrounding
medium of density $n_{0}$ forming a relativistic shock. A diminishingly small
fraction of electrons is accelerated by repeated diffusion across the shock 
producing a power-law energy
spectrum $N_{e}(\gamma_{e})\propto \gamma_{e}^{-p}$.
As the relativistic electrons spiral in the co-moving magnetic field we
expect to see synchrotron radiation.
A frequency $\nu_{m}$ is associated with the minimum electron
energy $\gamma_{e,min}$ produced by the acceleration process. There will
be some corresponding (very) high frequency $\nu^{*}$
associated with the maximum electron energy
$\gamma_{e,max}$. A cooling-break frequency, $\nu_{c}$,
corresponds to the electron energy for which the synchrotron lifetime is
comparable with the expansion time. The detailed form of the spectrum
expected is described by Sari, Piran and Narayan (1998).
At some early time $t_{0}$, $\nu_{m}=\nu_{c}$. For $t<t_{0}$ the electrons
cool rapidly so that $\nu_{m}>\nu_{c}$ and for $t>t_{0}$ they cool slowly
such that $\nu_{m}<\nu_{c}$. All the follow up observations
of GRB~030329 fall into the latter regime.
In the frequency range $\nu_{m}<\nu<\nu_{c}$ the
spectrum, $F(\nu)\propto\nu^{\beta}$, has an index $\beta$
directly related to the electron index $p$,
$\beta=-(p-1)/2$. For $\nu>\nu_{c}$
the spectrum is steeper, $\beta\sim-p/2$. Below the peak, $\nu<\nu_{m}$,
we expect
$\beta\sim+1/3$ with the index increasing further to $\beta\sim+2$,
at $\nu_{a}$ the
self-absorption break, as the optical depth increases at low frequency.

Because the fireball is expanding relativistically with Lorentz factor
$\Gamma$, the spectrum observed is blue-shifted. As the shock is arrested by
the surrounding medium $\Gamma$ decreases and the spectrum becomes
redder. At early times we only see a small fraction of the radiating
electrons because of relativistic beaming and only a small fraction
of the expanding shock is visible. As $\Gamma$ decreases the relativistic
beaming angle opens up and if the expansion is isotropic the
observed flux at the peak, $F_{m}$, remains constant while the
frequency of the peak changes as a power-law, $\nu_{m}\propto t^{-3/2}$.
The self-absorption break, $\nu_{a}$, remains constant but the cooling 
break also varies as a power law, $\nu_{c}\propto t^{-1/2}$. Wijers \&
Galama (1999) provide full expressions for the expected synchrotron spectrum
parameters in terms of energy per steradian
$\varepsilon_{52}=E_{52}/4\pi$, ambient interstellar density
$n_{0}$ cm$^{-3}$,
$X$ the hydrogen mass fraction of the ambient medium,
$p$ the electron power-law index,
$\chi_{p}$ and $\phi_{p}$ the dimensionless peak location and peak
flux which are functions of $p$,
$\epsilon_{e}$ the electron energy fraction (wrt the energy density of
nucleons), $\epsilon_{B}$ the magnetic energy density fraction
(again wrt the energy density of nucleons), the redshift $z$ of the GRB
and finally $h_{70}=H_{0}/70$ km s$^{-1}$ Mpc$^{-1}$.

If the relativistic outflow is collimated in a jet with open angle
$\theta_{j}$ then the evolution of the spectrum is modified
(Rhoads 1997, 1999; Sari, Piran \& Halpern 1999; Frail et al. 2001;
Panaitescu \& Kumar 2001b).
At time $t_{j}$, when $\Gamma^{-1}\approx \theta_{j}$, we
expect to see an achromatic break in the decay of the afterglow because
we can see the edge of the jet structure and because the material has
time to expand sideways.
After the break the decay of the peak flux, peak frequency, the cooling
break and the self-absorption break
are modified, $F_{m}\propto t^{-1}$, $\nu_{m}\propto t^{-2}$,
$\nu_{c}=const.$ and $\nu_{a}\propto t^{-1/5}$. The details of the
changes in the decay indices depends on the density profile of the 
surrounding medium and the time dependence of the jet spreading assumed.
The energy losses of the emitting electrons and the
observed spectral form is also expected to be
modified by inverse Compton scattering (Panaitescu \& Kumar 2001a,
Sari \& Esin 2001).

If $F_{m}\propto t^{\alpha_{f}}$, $\nu_{m}\propto t^{\alpha_{m}}$ and
$F(\nu)\propto \nu^{\beta_{0}}$
then the flux at a frequency $\nu_{m}<\nu<\nu_{c}$
has a temporal variation given by:
\begin{equation}
F_{\nu}\propto t^{\alpha_{f}-\alpha_{m}\beta_{0}}
\label{eq1}
\end{equation}
If $\nu_{c}\propto t^{\alpha_{c}}$ then for frequencies $\nu>\nu_{c}$
the spectral form is $F(\nu)\propto \nu^{\beta_{1}}$
and the temporal variation is 
$F_{\nu}\propto t^{\alpha_{f}-\alpha_{m}\beta_{0}-\alpha_{c}
(\beta_{1}-\beta_{0})}$. The change in spectral index across the
cooling break is $\beta_{1}-\beta_{0}=-p/2+(p-1)/2=-1/2$ so at
frequencies above the cooling break we have:
\begin{equation}
F_{\nu}\propto t^{\alpha_{f}-\alpha_{m}\beta_{0}+\alpha_{c}/2}
\label{eq2}
\end{equation}
Table \ref{tab1} lists the temporal indices of the peak flux
and break frequencies expected before $t_{0}$ and before and
after the jet break (Sari, Piran \& Halpern 1999).
\begin{table}
\centering
\caption{Temporal decay indices before $t_{0}$ and 
before and after the jet break $t_{j}$. Two
sets of values are given for $t<t_{0}$, the first line for an adiabatic
shock and the second line for the full radiative case.
$F_{m}\propto t^{\alpha_{f}}$,
$\nu_{a}\propto t^{\alpha_{a}}$,
$\nu_{m}\propto t^{\alpha_{m}}$,
$\nu_{c}\propto t^{\alpha_{c}}$.}
\begin{tabular}{@{}ccccc@{}}
             & $\alpha_{f}$ & $\alpha_{a}$ & $\alpha_{m}$ & $\alpha_{c}$ \\
\hline
$t<t_{0}$ A   & 0            & -1/2         & -3/2         & -1/2 \\
$t<t_{0}$ R   & -3/7         & -4/5         & -12/7        & -2/7 \\
$t<t_{j}$    & 0            & 0            & -3/2         & -1/2 \\
$t>t_{j}$    & -1           & -1/5         & -2           &  0   \\
\hline
\end{tabular}
\label{tab1}
\end{table}

\section{Multi-frequency data}

To calculate a reliable estimate of the physical parameters of a GRB
afterglow under the
fireball shock model we must measure $F_{m}$, $\nu_{m}$, $\nu_{c}$ and
$\nu_{a}$. However, this is not easy to do directly.
Most studies of GRB afterglows to date have concentrated on the interpretation
of light curves in a single waveband, in some cases augmented by sparse
multi-frequency data. Much attention has been given to the study of
decay breaks in light curves. These may correspond to a jet break
(if achromatic) or
a spectral break passing through a given pass band as the afterglow
reddens. In order to extrapolate the decay behaviour in a single
waveband we must have a reliable model for the decay, especially after
a jet break when the details depend on the structure of the surrounding
medium and the expansion at the edge of the jet.

The afterglow of GRB~030329 was particularly bright and has therefore
been subjected to extensive temporal and spectral observational coverage.
We have gathered together published photometric data from the afterglow
in the radio, sub-millimetre, IR, optical and X-ray 
energy bands. We have re-analysed the data from the XMM observations yielding
flux and spectral index estimates consistent with
Tiengo et al. 2003c.
Table \ref{tab2} summarises all the
X-ray measurements with the fluxes quoted in Jy at 1 keV. The RXTE results
are taken from Tiengo et al. (2003c).
\begin{table}
\centering
\caption{Summary of X-ray observations of GRB~030329. The fluxes and spectral
index $\beta$ are for a photon energy of 1 keV.}
\begin{tabular}{@{}ccccc@{}}
             & t days & Flux Jy & index $\beta$ \\
\hline
RXTE    & 0.215 & $(4.28\pm0.12)\times10^{-5}$ & $-1.17\pm0.04$ \\
RXTE    & 0.259 & $(3.41\pm0.19)\times10^{-5}$ & $-1.17\pm0.04$ \\
RXTE    & 1.26 & $(3.10\pm0.47)\times10^{-6}$ & $-0.80\pm0.3$ \\
XMM     & 37.3 & $(5.74\pm0.35)\times10^{-9}$ & $-1.08\pm0.10$ \\
XMM     & 61.0 & $(2.82\pm0.19)\times10^{-9}$ & $-1.15\pm0.12$ \\
\hline
\end{tabular}
\label{tab2}
\end{table}
The optical fluxes were calculated assuming a Galactic reddening correction
of E(B-V)=0.025 mag. (Burenin et al. 2003b, Matheson et al. 2003).

\begin{figure}
\centering
\includegraphics[height=8cm,angle=-90]{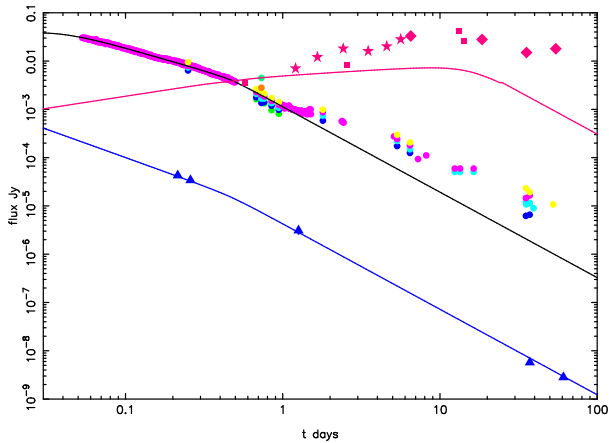}
\caption{The afterglow flux in X-rays (1.0 keV plotted in blue), optical (V, B, R and I plotted respectively as blue, green, red and yellow dots)
and radio (7.7 GHz diamond, 8.5 GHz square and 15.2 GHz star plotted in red).
The curves show the best fit model in X-ray (1 keV), optical (R-band) and
radio (8.5 GHz).
Data sources (all 2003):
X-ray - Marshall and Swank; Marshall, Markwardt and Swank;
Tiengo et al.;
Optical/IR -
Bikmaev et al.;
Burenin et al.;
Fitzgerald and Orosz;
Gorosabel et al.;
Ibrahimov et al.;
Lamb et al.;
Lee et al.;
Price et al.;
Simoncelli et al.;
Stanek et al.;
Susuki et al.;
Testa et al.;
Uemura et al.;
Zharikov et al.;
Sub-millimetre - Hoge et al.;
Radio -
Berger et al.;
Bloom et al.;
Finkelstein et al.;
Kuno et al.;
Pooley;
Rao et al.;
Trushkin et al.
}
\label{fig1}
\end{figure}

Fig. \ref{fig1} shows the evolution of the afterglow in 
the three main energy bands.
We have performed a multi-frequency fit using the fireball shock model
in a similar manner to Harrison et al. (2001),
Panaitescu \& Kumar (2001a,2002,2003),
Frail et al. (2003) and Yost et al. (2003). The coverage of the
present data for GRB~030329 is much greater in time and frequency than
for any previous burst and therefore the fitting is constrained at
a far higher level.
The solid lines in Fig. \ref{fig1}
show the best fit afterglow model which is described in detail below.
A clear temporal break is seen in the decay curves of each waveband.
Table \ref{tab3} lists the decay indices before ($\alpha_{0}$) and after
($\alpha_{1}$) the break time ($t_{b}$) estimated using simple power-law
fits.
\begin{table}
\centering
\caption{Power law decay indices before and after the break, 
and break times for the X-ray, optical
and radio light curves of the afterglow. Error ranges are $\pm1\sigma$
estimated from the least squares fitting procedure.}
\begin{tabular}{@{}cccccc@{}}
             & $\alpha_{0}$ & $\alpha_{1}$ & $t_{b}$ days \\
\hline
X-ray        & $-1.2\pm0.1$ & $-1.80\pm0.06$ & $0.57\pm0.1$ \\
Optical      & $-0.87\pm0.05$ & $-1.84\pm0.07$ & $0.52\pm0.05$ \\
Radio        & $+0.89\pm0.08$ & $-0.55\pm0.10$ & $9.86\pm0.06$ \\
\hline
\end{tabular}
\label{tab3}
\end{table}
The optical light curve suffers an initial break at $\sim0.5$ days
steepening to $\alpha_{1}=-1.84$ as
indicated in Table \ref{tab3} but at $\sim1$ day it flattens
off to $\alpha_{2}=-0.73$,
finally turning over again at $\sim3$ days to $\alpha_{3}=-1.35$.

\begin{figure}
\centering
\includegraphics[height=8cm,angle=-90]{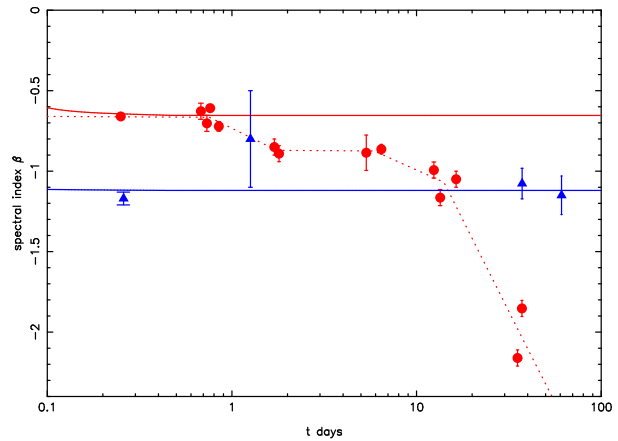}
\caption{The spectral indices of the afterglow in X-rays (1.0 keV plotted
as blue triangles) and
optical (plotted as red circles).
The solid lines show the evolution of the spectral index of the
optical and X-ray in the best fit model of the afterglow component.
The dotted line indicates the observed evolution of the optical spectral index.
Note that the earliest X-ray point is derived from the combination of
the first two RXTE observations (see Table \ref{tab2}).
Data sources (all 2003):
X-ray - Marshall and Swank; Marshall, Markwardt and Swank;
Tiengo et al.;
Optical -
Burenin et al.;
Fitzgerald and Orosz;
Ibrahimov et al.;
Lamb et al.;
Lee et al.;
Simoncelli et al.;
Testa et al.;
Zharikov et al.
}
\label{fig2}
\end{figure}

Optical spectroscopy indicates that between 1 and 10 days an optical
SN was emerging from the baseline afterglow of the GRB
(Stanek et al. 2003b). Fig. \ref{fig2} shows the measured values of the 
optical and X-ray spectral indices.
The optical BVRI index is $\beta_{0}=0.66\pm0.01$ at 0.25 days (Burenin et al. 
2003b)
and gradually reddens for $t>1.0$ days as the SN component starts to dominate.
The evolution of the optical spectrum is analysed in greater detail by
Matheson et al. (2003). Their data and analysis confirm the gradual
change in spectral index for $1<t<5$ days followed by a more dramatic
change for $t>6$ days as shown in Fig. \ref{fig2}.
The weighted mean of the four X-ray measurements is $\beta_{X}=1.13\pm0.06$.
The solid lines are the predicted indices from the best fit afterglow
model (see below) and the dotted line indicates the gradual change of the
optical index which starts at $\sim1$ day.

The difference between the optical and X-ray spectral indices for
$t<1.0$ days is $-0.47\pm0.06$ consistent with the expected change
across the cooling break frequency, $\nu_{c}$, $\beta_{1}-\beta_{0}=-0.5$.
Before $t_{b}=0.57$ days in the X-ray light curve the temporal
index is $\alpha_{0}=-1.2\pm0.1$ which is consistent with the
expected value before the jet break of
$-\alpha_{m}\beta_{0}+\alpha_{c}=-1.3$
calculated using equation \ref{eq2} if $\beta_{0}=0.7$.
Similarly the optical temporal index $\alpha_{0}=-0.87\pm0.05$
is consistent with the expected value $-\alpha_{m}\beta_{0}=-1.0$ from
equation \ref{eq1}. The slightly low optical value probably 
arises because the
curvature of the afterglow spectrum produces a smaller
value of $\beta_{0}$ near the peak.
The temporal decay indices of the X-ray and optical
light curves immediately after $t_{b}$ are very similar, $\alpha_{1}\sim-1.8$.
If we interpret $t_{b}$ as the jet break $t_{j}$ we can use equations
\ref{eq1} and \ref{eq2} to predict $\alpha_{1}=-2.4$ for both X-rays
($\nu>\nu_{c}$) and optical ($\nu<\nu_{c}$) since $\alpha_{c}=0$ for
$t>t_{j}$. Although the X-ray and optical light curves are decaying
at the same rate this rate is somewhat less than expected for $t>t_{j}$.
This maybe because $\alpha_{f}\sim0.5$ (see below) 
rather than $\alpha_{f}\sim-1.0$,
as listed in Table \ref{tab1}, or $\alpha_{m}$ did not change
from $-3/2$ to $-2$ as expected.

The optical data for $t<0.1$ days and $0.6<t<1.0$ days, the earliest
8.46 GHz measurement (Berger et al. 2003a), the earliest 1.288 GHz
measurement (Rao et al. 2003a,b) and all the X-ray data were
fitted using a $\chi^{2}$ statistic constructed from $\log F(\nu,t)$ and
the expected fireball shock spectrum including a
jet break.
The X-ray data are particularly important because they cover
a wide temporal window and are not expected to include flux from any
supernova component or the host galaxy.
The radio and optical data for $t>1.0$ days were excluded
from the fit because the rise in these two wavebands was probably
augmented by flux from the detected SN.
The optical data in the range
0.1-0.6 days were excluded because an optical bump in this
range resulted in a higher $\chi^{2}$ value. This small optical
bump is noticeable in the original plot of the optical data
presented by Uemura et al. (2003). A total of 207
data points gave $\chi^{2}=275$. With 8 free parameters in the model
there were 199 degrees of freedom and the $\chi^{2}_{\nu}=1.38$.
If the optical data in the range 0.1-0.6 days were included the reduced
Chi-squared increased, $\chi^{2}_{\nu}=2.64$.
The best fit parameter values and
estimated 90\% confidence limits are given in Table \ref{tab4}.
These parameter values are not significantly changed if we include
the optical data 0.1-0.6 days. It is remarkable the standard fireball
shock model is a good fit 
for all wavebands (X-ray, optical and radio) for $t<1$ day
and the agreement continues up to $t\sim60$ days in X-rays.
Furthermore the model provides an excellent fit to the measured
spectral indices in the optical ($t<1$ day) and X-rays as
indicated in Fig. \ref{fig2}.
\begin{table}
\centering
\caption{Best fit parameter values and 90\% confidence limits
for the parameters of fireball shock spectral model at $t=0.2$ days}
\begin{tabular}{@{}ccc@{}}
\hline
Fm Jy		&  0.0521   & 0.0507-0.0537 \\
$\nu_{a}$ Hz	&  $7.4\times10^{9}$ & $(5.3-10)\times10^{9}$ \\
$\nu_{m}$ Hz	&  $3.3\times10^{13}$ & $(3.1-3.5)\times10^{13}$ \\
$\nu_{c}$ Hz	&  $1.3\times10^{16}$ & $(0.9-1.8)\times10^{16}$ \\
p 		&  2.25		      & 2.24-2.26 \\
$t_{j}$ days	&  0.47               & 0.42-0.51 \\
$\alpha_{f}$    &  -0.48               & -(0.43-0.54) \\
$n_{sm}$	&  2.1                & 1.8-2.5    \\
\hline
\end{tabular}
\label{tab4}
\end{table}

The $n_{sm}$ parameter was used to smooth the broken power-law
sections in the way described by Beuermann et al. (1999).
This produces the
curvature required to fit the early optical light curve and is
responsible for the slight rise in the optical spectral index
at early epochs as shown in Fig. \ref{fig2}.
Explicitly, the model spectrum consists of four power-law sections
combined using a smoothing parameter:
\begin{equation}
F(\nu,t)=(F_{0}^{-n_{sm}}
+F_{1}^{-n_{sm}}
+F_{2}^{-n_{sm}}
+F_{3}^{-n_{sm}})^{-1/n_{sm}}
\end{equation}
where $F_{i}=k_{i}\nu^{\beta_{i}}$. The breaks between the sections
occur at $\nu_{a}$, $\nu_{m}$, $\nu_{c}$ which evolve with $t$
as indicated in 
Table \ref{tab1}. The order of the break frequencies depends on $t$
because they evolve at different rates.
The power-law indices for $t>t_{0}$ are given in section 2.
If the model is used for $t<t_{0}$ then we set
$\beta_{2}=-1/2$ but this does not effect the data fitting and
only applies when the model is used at very early times.
The normalisations $k_{i}$ are calculated to give
a continuous curve with a peak value $F_{m}$ (before smoothing).
The effect of the smoothing is only seen close to the breaks
as indicated in the model curves plotted in Fig. \ref{fig4}.

\begin{figure}
\centering
\includegraphics[height=8cm,angle=-90]{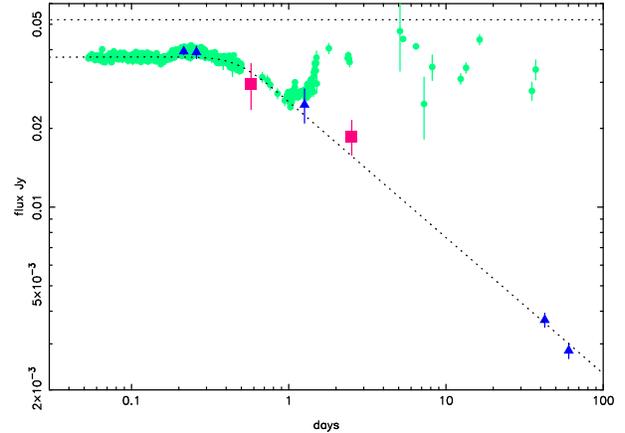}
\caption{Residuals about peak flux of the best fit afterglow.
Because we performed the fitting
in $\log F(\nu,t)$ the difference between data and the
model (calculated at the observed frequency)
is naturally expressed as a ratio. This ratio is shown multiplied
by the peak flux of the model 
($F_{m}$) so that the jet break in the model is apparent.
The horizontal dotted line indicates the peak flux
expected
without smoothing or the jet break. The dotted curve is the
peak flux with smoothing and including the jet break.
Optical R-band plotted as circles, Radio (7.7, 8.5 and 15.2 GHz)
squares, X-ray (1 keV) triangles.}
\label{fig3}
\end{figure}

\begin{figure}
\centering
\includegraphics[height=8cm,angle=-90]{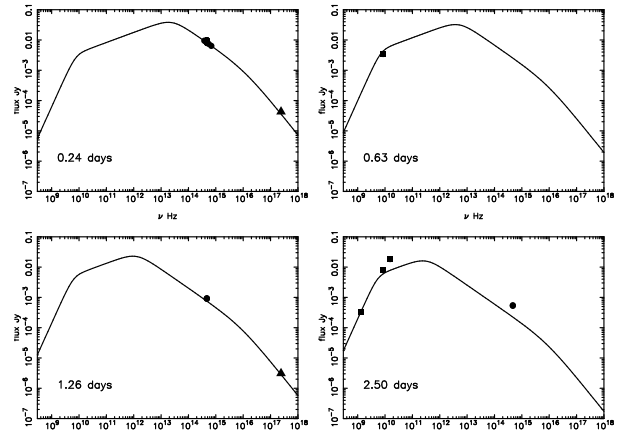}
\caption{The evolution of the broadband spectrum. The best fit continuum
is shown as a curve.
The measured points are X-ray (triangles), optical (circles)
and radio (squares).
Error bars were plotted but they are too small to
be seen at this scale. The lowest frequency radio data point (1.288 GHz)
in the bottom right panel is taken from Rao et al. (2003a,b) and
constrains the $\nu_{a}$ value in the model.}
\label{fig4}
\end{figure}

The temporal decay index of $F_{m}$ after the jet break, $\alpha_{f}=-0.48$, is
determined by the last three X-ray data points and the optical measurements
$0.6<t<1.0$ days.
This result is consistent with the analysis presented
by Tiengo et al. (2003c) but identifies the X-ray temporal decay
index $\alpha_{1}=-1.8$ with the decay in the peak flux of the
fireball spectrum.
Fig. \ref{fig3} shows the residuals about the peak flux ($F_{m}$) of the
best fit model.
The last three X-ray data values clearly define the peak decline
after the jet break.
Fig. \ref{fig3} also includes the optical data for $0.1<t<0.6$ and
$t>1.0$ which were excluded from the fitting procedure. The faint optical bump
with a peak at $\approx 0.25$ days and the dramatic rise in the optical
flux just after $t=1.0$ day are apparent.
Figure \ref{fig4} shows a series of four snap shots of the evolution of
the spectrum. In the lower panels at 1.26 and 2.5 days the optical and
higher frequency radio measurements are starting to lift away
from the underlying afterglow component. The absorption break $\nu_{a}$
is the least constrained parameter because it is only determined by
the radio measurements included at 0.63 and 2.5 days. All plots cover a flux 
range of $10^{-7}$ to $0.1$ Jansky and $3\times 10^{8}$ to $10^{18}$ Hz.
We attempted
to find an acceptable fit including the later optical and radio data.
It was possible to get a fit
without including a jet break but this predicted an X-ray flux
a factor of $\sim20$ greater than observed for $t>30$ days and
a factor of $\sim3$ smaller than the observed for $t<1$ day and,
furthermore, the
residuals for the optical data over the period $0.5<t<3.0$ days were
unacceptably large.
At the same time the radio measurements were poorly represented.
In particular the mean flux densities at 3.9, 7.7 and 11.2 GHz
from Trushkin et al. (2003) gave spectral indices which contradict the
expected trend.
Scintillation is expected to introduce variability in the observed
radio fluxes. For the Galactic coordinates $l_{II}=217$, $b_{II}=+60.7$
of GRB~030329 the rms variability expected at 8.5 GHz is $\sim30\%$
with a time scale of a few hours (Walker 1998) consistent with
the errors used at this frequency. The variability
will be larger at 1.288 GHz but the flux used in the fitting from
Rao et al. (2003a,b) for 31 March 2003 is the average from 8 hours
observation with 2-sigma error bars. We conclude that the
best fit of the afterglow model is not susceptible to scintillation.

\begin{figure}
\centering
\includegraphics[height=8cm]{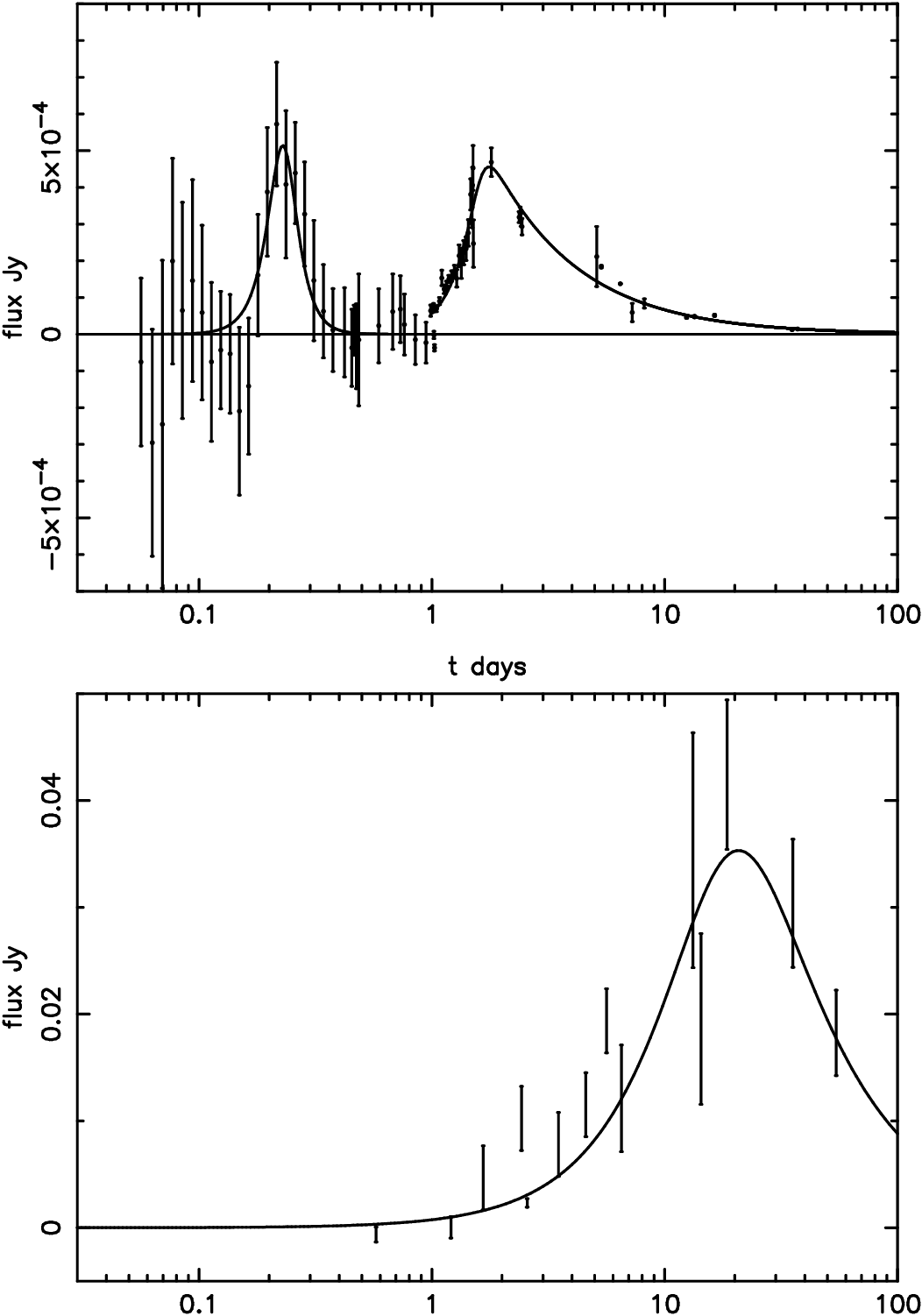}
\caption{The residual optical (R-Band) and radio
flux (7.7, 8.5 and 15.2 GHz)
after subtracting the best fit fireball shock model.
The early optical data points ($t<2.0$ days)
have been binned to improve the statistics. The radio fluxes have not
been corrected to a nominal frequency.
The curves are simple power-law fits to the rise and fall of each peak.
The first two radio data points (at 8.46 GHz) are consistent with the standard
fireball model.}
\label{fig5}
\end{figure}

Figure \ref{fig5} shows the residual optical and radio flux seen over
and above the best fit fireball shock model. The model is an excellent
fit to the optical data $t<0.2$ days and $0.6<t<1.0$ days and the
very early radio points. However there is highly significant additional
flux seen in both the radio and optical which does not fit the 
standard model (see also Tiengo et al. 2003c).

\section{Physical parameters}

We have used the formulation of Wijers \& Galama (1999) to calculate
the physical parameters of the afterglow. We assumed a hydrogen mass
fraction $X=0.7$ and used the $p=2.25$ value to estimate
$\phi_{p}=0.62$ and $\chi_{p}=0.56$ on the plots provided by
Wijers \& Galama (1999). The results are listed in 
Table \ref{tab5}.
\begin{table}
\centering
\caption{Physical parameters derived from the best fit afterglow model
together with 90\% confidence ranges.}
\begin{tabular}{@{}ccc@{}}
\hline
$\varepsilon_{52}$ 		&  6.8  & 5.1-9.0 \\
$n_{0}$ cm$^{-3}$	&  1.0   &  0.26-4.1\\
$\epsilon_{e}$	&  0.24 & 0.18-0.31 \\
$\epsilon_{B}$	&  0.0017 & 0.0007-0.0038 \\
p 		&  2.25		      & 2.24-2.26 \\
$\theta_{j}$ degrees & 3.0  &  2.4-3.9\\
$E_{50}$ (ergs in jet)    & 1.9 &  1.7-2.1 \\
\hline
\end{tabular}
\label{tab5}
\end{table}
The $\theta_{j}$ and $E_{50}$ values were derived from the formula
given by Sari, Piran and Halpern (1999) which specifies the jet break
as the time when lateral adiabatic expansion of the jet becomes
apparent.
The ranges of physical parameters given in Table \ref{tab5}
were estimated from the 90\% confidence ranges of the fit parameters
given in Table \ref{tab4}. In fact the ranges are dominated by the
uncertainty in $\nu_{a}$.

The index $\alpha_{f}=-0.48$ for the temporal evolution of $F_{m}$ for
$t>t_{j}$ is the only afterglow parameter
which does not conform to the standard model. If the expansion
after the jet break is adiabatic the we expect $\alpha_{f}\sim-1$.
This result suggests that either the edge of the jet is not well defined
(the jet has a broad smooth profile rather than a sharp edge,
Panaitescu \& Kumar 2003) or
something is confining the expected expansion.

Using the afterglow model we can extrapolate back to the epoch of the
GRB. From the best fit values of $\nu_{m}$ and $\nu_{c}$ we find
$t_{0}=5\times10^{-4}$ days (43 seconds) which is only $\sim18$ secs
after the burst faded. Using the
full radiative evolution values for $t<t_{0}$ from Table \ref{tab1} we 
get a total fluence in the 25 second burst of $1.1\times10^{-4}$ ergs 
cm$^{-2}$ which is remarkably close to the measured estimate of 
$1.2\times10^{-4}$ ergs cm$^{-2}$. The $F_{m}$ value must have remained
constant from 25 seconds to the break at 0.47 days as expected
from the standard model. The afterglow
model gives $\varepsilon\approx 6.8 \times 10^{52}$ ergs per steradian
for the shock,
while the fluence from the GRB was $\varepsilon_{52\gamma}\approx0.07$.
Hence we estimate that
$\sim1\%$ of the explosion energy was radiated as $\gamma$-rays in the
GRB.

The spectral analysis used to derive the physical parameters
in Table \ref{tab5} assumes that the afterglow spectrum is
dominated by synchrotron emission, ignoring any contribution from
self-Compton emission from the relativistic shock. Inverse Compton (IC)
scattering is expected to modify the high-energy tail of the spectrum
producing a characteristic bump in the X-ray and $\gamma$-ray energy
bands. Using the formulation of Sari \& Esin (2001)
we can use the physical shock
parameters in Table \ref{tab5} to estimate the photon energy
at which the synchrotron and IC component fluxes are the
same. The result is $>50$ keV for all epochs.
Therefore the bulk of any IC emission is expected to be above the RXTE and XMM
energy band and hence unobserved.
IC cooling is expected to be important at early times
in the afterglow evolution even if it is not observed directly.
In the fast cooling stages IC scattering increases the radiation losses
by a factor $\sim(\epsilon_{e}/\epsilon_{B})^{1/2}\approx12$ thereby
delaying the transition to the slow cooling regime to $t_{0}\approx0.15$
hours. However this is still well before the first optical observations
of GRB~030329 were made.
IC cooling continues to be important for $t>t_{0}$ while the fireball
(or jet) is relativistic. After the system has proceeded through the
jet break and ceases to be highly 
relativistic the efficiency of IC cooling drops rapidly and ceases to
be important. Since we observe a break which is probably the jet break
at $t\approx0.5$ days we can be confident that our results are
not compromised by IC emission.

\section{Origin of the excess flux}

There are two optical bumps in the residuals plot Fig. \ref{fig5}.
The solid lines in the plots are simple power fits used to estimate
the peak positions and integrated fluxes.
The first R-band peak has a maximum at 0.23 days, a FWHM of 0.06 days 
(5\% peak flux range 0.14 to 0.35 days), 
and an integrated flux of 5.5 Jy secs corresponding to a total
energy of
$\approx1.4\times10^{48}$ ergs in the optical waveband
($\Delta\nu\approx4\times10^{14}$ Hz).
The rise time is
only $\sim0.09$ days and the maximum flux is $\sim6\times10^{-4}$ Jy
corresponding to $R_{mag}\approx16.8$.
The second R-band peak is a much longer. It has a maximum at 1.7 days, 
a FWHM of 2.0 days (5\% peak flux range 0.9 to $\sim$28 days), 
and an integrated flux of 234 Jy secs corresponding
to a total of $\approx5.9\times10^{49}$ ergs in the optical waveband.
Again, the rise time of $\sim0.8$ days is remarkably short,
the maximum flux is $\sim4.5\times10^{-4}$ Jy corresponding to
$R_{mag}\approx17.1$.
It is interesting that the spectral index of the optical spectrum
starts to redden at $\sim1$ day (see Fig. \ref{fig2}) at the
onset of this second R-band peak. Although the maximum flux
for the two optical peaks is similar, because the afterglow is
decaying rapidly the first R-band peak represents
only a small increase over and above the underlying afterglow,
$\sim5\%$, while the second peak corresponds to a large increase of $\sim100\%$.

A single broad radio peak (7.7, 8.5 and 15.2 GHz data points)
is also shown plotted in Fig. \ref{fig5}.
It has a maximum of $\sim0.035$ Jy (at $\sim8$ GHz) at $\sim20$ days
corresponding to the time of emergence of the full
SN spectrum, Hjorth et al. (2003). The peak luminosity is
$\sim3\times10^{31}$ ergs s$^{-1}$ Hz$^{-1}$. The properties of the first
and second R-band peaks and the radio peak are summarised in Table \ref{tab6}.
\begin{table}
\centering
\caption{Properties of the flux peaks in excess of the best fit
afterglow model.}
\begin{tabular}{@{}cccc@{}}
	& first R-band & second R-band & radio $\sim8$ GHz \\
\hline
$t_{start}$ days & 0.14 & 0.9 & 1.0 \\
$t_{max}$ days & 0.23  & 1.7 & 20 \\
$t_{end}$ days & 0.35  & $\sim28$  & $>100$ \\
FWHM days & 0.06 & 2.0 & $\sim38$ \\
rise time days & 0.09  & 0.8 & $\sim17$ \\
peak flux Jy   & $6\times10^{-4}$ & $4.5\times10^{-4}$ & 0.035 \\
peak $R_{mag}$ & 16.8 & 17.1 & - \\
optical ergs   & $\sim1.4\times10^{48}$ & $\sim5.9\times10^{49}$ & - \\
\hline
\end{tabular}
\label{tab6}
\end{table}

There are a number of possible explanations for the excess flux seen
over and above the standard fireball shock. We will discuss them
under three headings; Afterglow effects - modifications to the
standard afterglow shock model;
Dust echoes - scattering of afterglow
radiation back into our line of sight;
SN/Hypernova - optical and radio flux from a stellar explosion.
A key feature of the observed residuals is that they
are not seen in X-rays and they are not the same in the optical and radio.
The first R-band peak is a short flash while the second R-band peak
is a longer outburst and the radio peak reaches a maximum at a much later
time than the second R-band peak.

\subsection{Afterglow effects}

If the ISM about the progenitor were inhomogeneous (varying density $n_{0}$)
then we might expect
to see rapid and noticeable deviations from a smooth power-law decay.
Such an effect was considered by
Wang \& Loeb (2000) and has been used to explain variations seen
in several afterglows (e.g. 
GRB~011211, Holland et al, 2002; Jakobsson et al, 2003; GRB
021004, Lazzati et al, 2003; Holland et al, 2003).
For $t<t_{j}$ (Nakar, Piran and Granot 2003) and below the cooling break
($\nu<\nu_{c}$, usually the optical regime) the flux is expected
to be $F_{\nu}\propto n^{(p+1)/4}$ whilst for $\nu>\nu_{c}$ (the X-ray regime)
the flux is only weakly dependent on density. Therefore the first
peak in the R-band flux, which corresponds to a very modest increase
$\Delta F_{\nu}\sim5\%$, could easily be due to a small increase
in ambient density. However, for $t>t_{j}$ the opposite is true
(Nakar, Piran and Granot 2002). Here we expect a change in density
to increase the X-ray flux more than the optical. It is therefore most
unlikely that the second R-band peak, which corresponds to an
increase of the optical afterglow flux by $\sim100\%$ and is
not accompanied by a similar large increase in the X-ray flux, is caused by
a rise in ambient density.

Refreshed shocks due to sustained activity in
the central engine have been used to explain
the observed lightcurve of GRB~021004, which was unusually flat for the
first few hours, and shows correlated X-ray and optical changes
(Fox et al., 2003).
This model has also been invoked for GRB~030329 (Granot, Nakar and Piran 2003).
Refreshed shocks should be seen as correlated outbursts
in the X-ray and optical decays although the fractional increase seen in
optical and X-ray is not expected to be the same
(Sari \& M\`{e}sz\`{a}ros 2000). At the same time we might also
expect to see a change in the X-ray and optical spectra.
Because of the sparse X-ray coverage we cannot rule out the possibility
that the optical excess arises from refreshed shocks and any associated
small increase in the X-ray flux was unfortunately missed. However, the
fit of the standard model to the available X-ray data is very good
and we see no change in X-ray spectral index.
We think it is unlikely that all the excess optical flux arises from
refreshed shocks. There is variability in the excess flux
of the second peak and at least three more re-brightenings were identified
by Granot, Nakar and Piran (2003) at $t\sim2.6$ days, $t\sim3.3$ days and
$t\sim5.3$ days. They favour refreshed shocks as the most likely
explanation for these features. However, analysis of millimeter observations
presented by Sheth et al. (2003) shows that there is no
short-lived millimeter emission associated with bumps in the optical
light curve indicating that central engine of the GRB was not injecting
energy on these timescales.

A further possibility is that we are observing lateral structure 
within the jet rather than axial structure imposed by the ambient medium.
This possibility is investigated by Granot \& Kumar (2003),
Kumar \& Granot (2003) and Salmonson (2003). 
However such structure is expected to
modify the overall evolution of the light curve producing, for
example, a late temporal break in the afterglow decay and is not
expected to produce rapid variations or bursts.

We conclude that at least the second optical bump
is not due to afterglow effects.

\subsection{Dust Echoes and Scattering}

Dust echos were considered to be a
possible explanation of late time bumps
in the light curves of some GRBs 
(Waxman \& Draine, 2000). In this scenario dust close to the 
burst is sublimed by the prompt X-ray and EUV radiation, whilst dust
beyond the sublimation radius absorbs the energy and then re-radiates
it at longer wavelengths. Bumps or flares are expected to occur at late
times after the GRB (the the exact times being a function of the
sublimation radius of the burst, and of course the assumed beaming
geometry). 
Because dust sublimes at $\sim 2000$K the bumps
are necessarily very red and are unlikely to have
significant flux in the optical band. We do not consider them
a viable explanation of the bumps in GRB~030329.

Esin \& Blandford (2000), and subsequently Reichart (2001), take a 
different approach in which the scattering of the initial optical flash by the 
dust is responsible for a bump in the afterglow light curve. Reichart
(2001) finds that for values of $\theta_j \leq 5$ degrees,
and small sublimation radii the
echo can reach maximum in less than one day, consistent with the first
R-band peak.
If the efficiency of the reflection/scattering process is $\sim0.1$
it is unlikely that these models can explain the second R-band
peak which has a very large integrated energy implying a
huge initial burst.

\subsection{SN/Hypernova}

At $z=0.1685$ GRB~030329 provides a unique opportunity to probe the
early behaviour of any associated SN. 
There is overwhelming evidence that the spectroscopic signature
of a SN Type Ic forms part of the optical afterglow of GRB~030329 (Hjorth
et al. 2003). This supernova has been designated SN2003dh
(Stanek et al. 2003a, Garnavich et al. 2003). The expansion
velocity measured at 10 days was $36,000\pm3,000$ km s$^{-1}$, larger
than any previously know SN and it is therefore worthy of the title 
hypernova. Crude spectroscopic dating concludes
that the SN was coincident with GRB~030329 $\pm2$ days
(Hjorth et al. 2003). Using the spectroscopic evidence alone Matheson et al.
(2003) conclude that SN2003dh emerges at $t>6$ days.
The photometric evidence (after subtraction of the standard afterglow
component) and the onset of the changes in optical spectrum for
$t>1.0$ days (see Fig. \ref{fig2})
indicate that the optical flux associated with SN2003dh is detected earlier
than this.

Any premaximum display of variability in the SN lightcurve
including the shock breakout and structure introduced
as the shock propagates through the layered structure of the star
may be visible. Indeed, the short first R-band peak followed by the
extended second R-band peak
shown in Fig. \ref{fig5} are very similar to the model light curves of
Type Ib SN models shown by Ensman \& Woosley (1998), although
the time scale is a factor of 6-10 faster. Models of Type Ic SNe by
Nakamura et al. (2001) indicate that the peak of the luminosity
for such events should occur in the range 10-15 days. 
The peak of the main outburst in Fig. \ref{fig5} is at $t\approx1.7$ days,
much earlier than predicted from the models.
SN1998bw peaked at $M_{V}=-19.35$, $M_{R}=-19.36$ (Galama et al. 1998).
Scaling this to z=0.1685 gives $R_{mag}=20.22$, therefore the peak of the long
optical outburst in Fig. \ref{fig5} $R_{mag}=17.1$ is $\sim18$ times brighter
than SN1998bw, having $M_{V}\approx-22.5$.

The SN 1b models of Ensman \& Woosley (1988) produce a main peak FWHM of
25 days as observed for SN1983N (the only event for which
good data are available), which is only
possible for the lower mass models
(Wolf-Rayet = 4-6 M$_{\odot}$, orig mass = 15-20 M$_{\odot}$).
More massive models create broader peaks, as do less powerful explosions.
A smaller ejected mass gives a lower duration
because the radioactive luminosity diffuses out quicker.
However a small ejected mass results in a dim SN since it is powered
by the ejected Ni.
In modelling SN 1998bw Nakamura et al. (2001) show that the
main peak width is proportional to the ejected mass
$M_{ej}^{3/4}$ and the explosion energy $E^{-1/4}$, while the
velocities scale as $M_{ej}^{-1/2}$ and $E^{1/2}$. These relationships
point to either a low ejected mass or a high powered explosion in GRB~030329
or both. The FWHM of
1998bw was $\sim29$ days and the ejection velocity at 10 days is $\sim26,000$
km s$^{-1}$. This is fit with $E=5\times10^{52}$ ergs and 
$M_{ej}=10$ M$_{\odot}$ by
Nakamura et al. (2001). Scaling from this we estimate the energy and mass of
SN2003dh as $E=5\times10^{51}$ ergs and $M_{ej}=0.6$ $M_{\odot}$. This
is a significantly lower ejected mass than considered by Ensman \& Woosley
(their values were based on there being a He envelope present).

Sheth et al. (2003) and Berger et al. (2003b) suggest that the observed
excess radio and millimeter flux (and by implication excess optical flux) from
GRB~030329 arises from a second slower and wider jet-like outflow.
The shock in such an outflow would generate a spectrum similar to the standard
model illustrated in Fig. \ref{fig4} but with the absorption break and peak
shifted to much lower frequencies. This could account for the broad radio
and optical excess but would not generate a significant X-ray flux,
consistent with the analysis presented above. It might also explain the
change in the optical spectral index which starts at t=1 day.
However it would not account for the very fast optical flux increase
seen at the onset of the second optical peak. This would require a
large extra injection of energy at t=1 day, similar in kind 
to a refreshed shock,
rather than a slower, wider outflow launched at t=0.
The duration and timing of the radio peak are very similar to known RSNe.
However, the peak luminosity of
$\sim3\times10^{31}$ ergs s$^{-1}$ Hz$^{-1}$ is a factor of $10^{4}$ higher
than typical Type Ib/c RSNe (Weiler et al. 2002) but very similar
to the radio afterglows seen for GRB~991208,
$\sim2\times10^{31}$ ergs s$^{-1}$ Hz$^{-1}$, and for GRB~970508, 
$\sim9\times10^{30}$ ergs s$^{-1}$ Hz$^{-1}$ 
(Frail, Waxman \& Kulkarni 2000) at 5 GHz. This, again,
clearly indicates that the hypernova in GRB~030329 was indeed a powerful event.

The combination of a rapid first peak and a very bright second peak is
not predicted by standard models of SNe. The second peak is
too bright and reaches peak too quickly to be entirely
fuelled by the radioactive decay
of a large mass of $^{56}$Ni (Woosley \& Heger 2003).
If the excess flux is directly
associated with the hypernova then it arises from the injection of
energy from some other source.
Both the optical and radio maxima were very bright compared with
previously observed SNe/hypernovae. This may be due to the geometry
of the explosion rather than just a very large explosion energy. The
ejecta could be beamed and/or the visible envelope of the expanding
photosphere might be flattened. Since the hypernova was concurrent
with a GRB this seems reasonable although we know of no detailed
models which address such a scenario.


\section{Conclusion}

The high quality and extensive coverage of data available from GRB~030329 has
enabled us to calculate the physical parameters of the event
with high precision and has revealed important details about
the concurrent hypernova SN2003dh.

The GRB released an energy of $\sim1.9\times10^{50}$ ergs into
a jet with open angle $\sim3$ degrees and about $1\%$ of this energy
was radiated away as $\gamma$ rays. The jet propagated in a medium
with $n_{0}=1$ cm$^{-3}$ and the shock generated an relativistic
electron spectrum
with index $p=2.25$. The electron and magnetic field energy densities
in the shock were $\epsilon_{e}=0.24$ and $\epsilon_{B}=0.002$.
A jet break occurred at $0.5$ days after which the peak flux
of the afterglow, $F_{m}$,
decayed with $\alpha_{f}=-0.5$.

It is likely that the lightcurves shown in Fig. \ref{fig5} are directly
associated with the hypernova SN2003dh. This event had the 
spectroscopic signature of a SN 1c by 8-10 days but the fast rise
and decline in the optical, the very bright optical peak and the large
radio excess are not standard SN 1c properties.
The excess optical and radio flux probably arise from 
an afterglow powered by a central engine such as the disk wind component
proposed by Woosley \& Heger (2003) but injection of energy at $t\approx1$ day
is required to explain the sudden emergence of the second optical peak.

\section{Acknowledgements}
We thank Makoto Uemura for supplying the early optical data from
Uemura et al. (2003) promptly following publication.
We also thank the ESA XMM-Newton team for their support of GRB science in
scheduling the XMM target of opportunity observations.
AL and KLP acknowledge the support of PPARC studentships.

\end{document}